\documentclass[prd,email,preprint,showkeys,preprintnumbers,amsmath,amssymb,nofootinbib]{revtex4}
\usepackage{graphicx}
\usepackage{amssymb}
\usepackage{epsfig}
\usepackage{dsfont}
\usepackage{amssymb,amsmath}
\usepackage{eufrak}
\usepackage{euscript}
\usepackage{latexsym}
\usepackage[latin1]{inputenc}
\usepackage{pstricks}

\newcommand{\be}{\begin{equation}}
\newcommand{\ee}{\end{equation}}
\newcommand{\ba}{\begin{eqnarray}}
\newcommand{\ea}{\end{eqnarray}}
\newcommand{\p}{\partial}

\skip\footins 20pt plus4pt minus4pt
\begin{document}

\title{\LARGE\textbf{Path integral formalism in a Lorentz \\ invariant noncommutative space}}

\author{Mario J. Neves$^a$}
\email{mariojr@ufrrj.br}
\author{Everton M. C. Abreu$^{a,b,c}$}
\email{evertonabreu@ufrrj.br}

\affiliation{${}^{a}$Grupo de F\' isica Te\'orica e Matem\'atica F\' isica, Departamento de F\'{\i}sica,
Universidade Federal Rural do Rio de Janeiro\\
BR 465-07, 23890-971, Serop\'edica, Rio de Janeiro, Brazil\\
${}^{b}$LAFEX, Centro Brasileiro de Pesquisas F\' isicas (CBPF), Rua Xavier Sigaud 150,\\
Urca, 22290-180, RJ, Brazil\\
${}^{c}$Departamento de F\'{\i}sica, ICE, Universidade Federal de Juiz de Fora,\\
36036-330, Juiz de Fora, MG, Brazil\\
\bigskip
\today\\}




\begin{abstract}


{\noindent We introduced a new formulation for the path integral formalism for a noncommutative (NC) quantum mechanics
defined in the recently developed Doplicher-Fredenhagen-Roberts-Amorim (DFRA) NC framework that can be considered an alternative framework for the NC spacetime of the early Universe.
The operators formalism was revisited and we apply its properties to
obtain a NC transition amplitude representation. Two DFRA's systems were discussed, the NC free
particle and NC harmonic oscillator. Some temperature concepts in this NC space are also considered.  
The extension to NC DFRA quantum field theory
is straightforward and we apply it to a massive scalar field.  We construct the generating functional
and the effective action to give rise one-particle-irreducible diagrams. As an example, we set the basis for a 
$n\;(n\geq 3)$ self-interaction $\phi^{n}$ to obtain the correction of the perturbation theory to the
propagator and vertex of this model.  The main concept that we would like to emphasize from the outset is that 
the formalism demonstrated here will not be constructed introducing a NC parameter in the system, as usual.  
It will be generated naturally from an already NC space.  In this extra dimensional NC space, we presented also the idea of dimensional reduction to recover commutativity.} \\ \vspace{0.7cm}

\end{abstract}

\maketitle

\pagestyle{myheadings}
\markright{Path integral formalism in a Lorentz invariant noncommutative space}

\section{Introduction}
\renewcommand{\theequation}{1.\arabic{equation}}
\setcounter{equation}{0}

The first published work concerning a noncommutative (NC) concept of spacetime was carried out in $1947$ by Snyder in his seminal paper \cite{snyder47}.
The need to control the ultraviolet divergences in quantum field theory (QFT) was the first motivation
to consider a NC spacetime. 

The main NC idea is that the spacetime coordinates $x^{\mu}\;(\mu=0,1,2,3)$
are promoted to operators in order to satisfy the basic commutation relation
\begin{eqnarray} \label{xmuxnu}
\left[\,{\mathbf x}^\mu\,,\,{\mathbf x}^\nu\,\right]\,=\,i\,\alpha\theta^{\mu\nu}\,\,,
\end{eqnarray}
where $\theta^{\mu\nu}$ is an antisymmetric constant matrix, and $\alpha$ is a length scale.
The alternative would be to construct a discrete spacetime with a NC algebra.
Consequently, the coordinates operators are quantum observable that satisfy the uncertain relation
\begin{eqnarray}\label{uncertainxmu}
\Delta {\bf x}^{\mu} \Delta {\bf x}^{\nu} \simeq \lambda \theta^{\mu\nu} \; ,
\end{eqnarray}
it leads to the interpretation that noncommutativity of spacetime must emerge in
a fundamental length scale $\alpha$, the {\it Planck scale}, for example.

However, Yang \cite{yang47}, a little time later, demonstrated that Snyder's hopes in cutting off the infinities in QFT
were not obtained by noncommutativity. This fact doomed Snyder's NC theory to years of ostracism.
After the important result that the algebra obtained with a string theory embedded in a magnetic
background is NC, a new perspective concerning noncommutativity was rekindle \cite{seibergwitten99}.
Nowadays the NC quantum field theory (NCQFT) is one of the most investigated subjects about the description
of a physics at a fundamental length scale of quantum gravity \cite{QG}.


The most popular noncommutativity formalism consider $\theta^{\mu\nu}$
as a constant matrix, as we said before. Although it maintains the translational invariance,
the Lorentz symmetry is not preserved \cite{Szabo03}.
To heal this ``disease" a recent approach was introduced
by Doplicher, Fredenhagen and Roberts (DFR) \cite{DFR}. It considers $\theta^{\mu\nu}$ as an ordinary
coordinate of the system in which the Lorentz symmetry is preserved.
Recently, it has emerged the idea \cite{Morita} of constructing an extension of the DFR spacetime
introducing the conjugate canonical momenta associated with
$\theta^{\mu\nu}$ \cite{Amorim1} (for a review the reader can see \cite{amo}).
This extended NC spacetime has ten dimensions:
four relative to Minkowski spacetime and six relative to $\theta$-space.
This new framework is characterized by a field theory constructed in a spacetime with extra-dimensions $(4+6)$,
and which does not need necessarily the presence of a length scale $\alpha$ localized into the six dimensions of the $\theta$-space, where $\theta^{\mu\nu}$ has dimension of length-square.
Besides the Lorentz invariance keeps maintained, obviously we hope that causality aspects in 
QFT in this $\left(x+\theta\right)$-spacetime must be preserved too \cite{EMCAbreuMJNeves2011}.

By following this conjecture, a new version of NC quantum mechanics (NCQM) was introduced.  In this formalism 
not only the coordinates ${\mathbf x}^\mu$ and their canonical momenta ${\mathbf p}_\mu$ are
considered as operators in a Hilbert space ${\cal H}$, but the objects of noncommutativity $\theta^{\mu\nu}$
also have their canonical conjugate momenta $\pi_{\mu\nu}$ \cite{Amorim1,Amorim4,Amorim5,Amorim2}.
All these operators belong to the same algebra and have the same hierarchical level, introducing a minimal canonical
extension of DFR algebra, the so-called Doplicher-Fredenhagen-Roberts-Amorim (DFRA) formalism. 

If the $\theta$ parameter is treated as a constant matrix, the non relativistic theory is not invariant under rotation symmetry.
This enlargement of the usual set of Hilbert space operators allows the theory to be invariant under the rotation
group $SO(D)$, as showed in detail in \cite{Amorim1,Amorim2}, where the treatment is a non relativistic one.
Rotation invariance in a non relativistic theory is the main ingredient if one intends to describe any physical system in a
consistent way.  In the first papers that treated the DFRA formalism, the main motivation, as we said before, was to construct a NC standard QM.

In this paper we introduce the NC path integral version for this NCQM defined in the $(x+\theta)$ DFRA space.  Our objective is to expand the boundaries of the NC path integral (NCPI) current literature.  Trying to put the reader updated the NCPI works we will sum up the main papers in this subject.  However, we will not follow any chronological order.  The sequence follows as the works came to our knowledge.  

In \cite{1} the authors constructed the NC formulation of PI suggesting that NCQM should be developed in the Hilbert space of Hilbert-Schmidt operators acting on classical configuration space and the concept of extended and structured objects for NCPI was explored in \cite{2}.  In \cite{3}, the NCPI was formulated through coherent states, which were the eigenstates of complex combinations of position operators.  In \cite{4} the NCPI formulation used the above proposed formulation of NCQM as a quantum system on Hilbert-Schmidt space using coherent spaces.  Coherent states were also used in \cite{15a}.  In \cite{4} the action computed for a particle moving in a NC plane with an arbitrary potential was nonlocal in time.  This nonlocallity was removed introducing an auxiliary field.  The limit $\theta \rightarrow 0$ recovered the free particle and harmonic oscillator propagators.  In \cite{5} the NCPI approach used the ideas of Connes' NC spaces. The spectral action concept was used in \cite{6}.  In \cite{7} the NC $\theta$-parameter was introduced in PI formulation of the transition amplitude via canonical noncommutativity.  The $\theta$-parameter was also introduced in \cite{8}, through the Bopp shift in the harmonic oscillator and its partition function at a finite temperature was calculated.  A two-dimensional sigma model at finite temperature was accomplished in \cite{9}.  A PI formula for the $*$-product of functions of $x$ and $\theta$ was proposed in \cite{10} where $\theta$ is a Grassmann variable (Majorana spinnor in $n$ spacetime dimensions) closing a supersymmetric NC algebra.  To derive the PI formula, the Batalin-Vilkovisky formalism was used.  A PI form was utilized in \cite{11} to construct a map between commutative and NC fields.  And the Schwinger's action principle was used in \cite{12} to develop a PI for NC spacetime theory.  The Bopp shift was also used in \cite{13} to formulate a quadratic Hamiltonian and an effective quadratic Lagrangian to construct a NCPI to compute NC probability amplitudes.  The same authors analyzed variations of NC algebra to introduce NCPI in \cite{14}.  A NCPI analysis was used for spinning and spinless particles in \cite{15}.  And PI's for the NC complex scalar field theory with self-interaction was accomplished in \cite{15}.

Hence, in this work we follow a completely different direction of all this papers.  In an already nine dimensional NC space we constructed a PI.  To recover the commutativity we used a different way also.  We promoted a dimensional reduction where the extra three NC dimensions were eliminated.  Although it is a simple procedure, as we will see, its concept is new in the NC literature where the $\theta \rightarrow 0$ limit is used (as it should be) by the huge majority of NC works.

The paper is organized as: the next section is dedicated (for self-containment of this work) to a review of the basics of QM in this NC DFRA framework, namely, the DFRA algebra.  We discussed the natural extension to NC quantum field theory (QFT).   In the third section we apply the basics of DFRA NCQM to construct a new NC path integral formalism.  Notice that the word new here is justified because this construction is accomplished within a NC six dimensional space, as we will explain.  In other words, it will not be construct ``introducing" a NC parameter.  The formalism will be generated naturally from an essentially NC space.  In the fourth section we apply the NC path integral to the free particle and harmonic oscillator.  The fifth section is dedicated to the extension of this formalism to NCQFT of a massive
scalar field.  We obtained the perturbative series via generating functional and $n$-points Green functions
in the space $x+\theta$. The effective action as the generating functional of the 1PI (one-particle-irreducible)
Green functions is presented too. 
To finish, we discussed the results obtained and we made the final remarks and conclusions.




\section{Quantum mechanics and field theory in the $DFRA$ NC spacetime}
\renewcommand{\theequation}{2.\arabic{equation}}
\setcounter{equation}{0}

From now on in this section  we will follow the main steps published in \cite{Amorim1,Amorim4,Amorim5,Amorim2}.  Namely,
we will revisit the basics of the NCQM defined in the DFRA space.
We assume that this space has $D \geq 2$ dimensions. The operators ${\bf x}_{i}$ $\left(i=1,2,...,D\right)$
and ${\bf p}_{i}$ are the position operator and its conjugated momentum. They satisfy the usual commutation relation
\begin{eqnarray}\label{xpcom}
\left[{\bf x}_{i},{\bf p}_{j}\right]=i\delta_{ij} \; ,
\end{eqnarray}
where we has adopted the Natural units $(\hbar=c=1)$ and $\alpha=1$.   Having said that, in NCQM we can rewrite the commutation relation between the position operator, that is
\begin{eqnarray}\label{xxcom}
\left[{\bf x}_{i},{\bf x}_{j}\right]=i\theta_{ij} \; ,
\end{eqnarray}
where $\theta_{ij}$ is an antisymmetric matrix. In DFR formalism $\theta_{ij}$ is considered as a space
coordinate and a position operator in $\theta$-space, of course.   Therefore this assumption leads us
to a space with coordinates of position $\left({\bf x}_{i},\theta_{ij}\right)$, in which $\theta_{ij}$
has $D(D-1)/2$ independent degrees of freedom. If $\theta_{ij}$ are coordinates, in addition to (\ref{xxcom})
it assumed that
\begin{equation}\label{xthetacomm}
\left[{\mathbf x}_{i},{\mathbf \theta}_{jk}\right] = 0
\hspace{0.4cm} \mbox{and} \hspace{0.4cm}
\left[{\bf \theta}_{ij},{\mathbf \theta}_{k\ell}\right] = 0 \; .
\end{equation}
Moreover exist the canonical conjugate momenta $\pi_{ij}$ associated with coordinates $\theta_{ij}$,
and they must satisfy the commutation relation
\begin{equation}\label{thetapicomm}
\left[\,\theta^{ij},\pi_{k\ell}\, \right] = i \delta^{ij}_{\,\,\,\,\,k\ell} \; ,
\end{equation}
where $\delta^{ij}_{\,\,\,\,\,k\ell}=\delta^{i}_{k}\delta^{j}_{\ell}-\delta^{i}_{\ell}\delta^{j}_{k}$.
In order to obtain consistency we can write that \cite{Amorim1}
\begin{equation}\label{ppicomm}
\left[{\mathbf p}_{i},{\mathbf p}_{j} \right] = 0
\hspace{0.2cm} , \hspace{0.2cm}
[{\mathbf p}_{i},{\mathbf \theta}_{jk}] = 0
\hspace{0.2cm} , \hspace{0.2cm}
[{\mathbf p}_{i},{\mathbf \pi}_{jk}] = 0 \; ,
\end{equation}
and this completes the DFRA algebra.

The Jacobi identity formed by the operators ${\mathbf x}_{i}$, ${\mathbf x}_{j}$
and ${\mathbf \pi}_{kl}$ leads to the nontrivial relation
\begin{equation}\label{nontrivialrel}
[[{\mathbf x}_{i},{\mathbf \pi}^{kl}],{\mathbf x}_{j}]- [[{\mathbf x}_{j},{\mathbf \pi}^{kl}],{\mathbf x}_{i}]
=- \delta^{ij}_{\;\;\;kl} \; ,
\end{equation}
which solution, not considering trivial terms, is given by
\begin{equation}\label{solutionnontrivial}
[{\mathbf x}_{i},{\mathbf \pi}^{jk}]=-{i\over 2}\delta_{il}^{\;\;\;jk}{\mathbf p}_{l} \,\,.
\end{equation}
It is possible to verify that the whole set of commutation relations listed above is indeed consistent
 under all possible Jacobi identities and the CCR algebras \cite{EMCAbreuMJNeves2011}. Expression (\ref{solutionnontrivial}) suggests that the shifted coordinate
operator \cite{Chaichan,Gamboa,Kokado,Kijanka,Calmet}
\begin{equation}\label{X}
{\mathbf X}_{i} := {\mathbf x}_{i}\,+\,{1\over 2}{\mathbf \theta}_{ij}{\mathbf p}^{j}\,\,,
\end{equation}
commutes with ${\mathbf \pi}_{kl}$.  The relation (\ref{X}) is also known as Bopp shift in the literature. 
The commutation relation (\ref{solutionnontrivial})
also commutes with ${\mathbf \theta}_{kl}$ and $ {\mathbf X}_{j} $, and satisfies a non trivial
commutation relation with ${\mathbf p}_{i}$ dependent objects, which could be derived from
\begin{equation}\label{Xpcomm}
[{\mathbf X}_{i},{\mathbf p}_{j}]=i\delta_{ij}
\hspace{0.6cm} \mbox{and} \hspace{0.6cm}
[{\mathbf X}_{i},{\mathbf X}_{j}]=0\,\,.
\end{equation}
So we see from these both equations that the shifted coordinated operator (\ref{X}) allows us to recover
the commutativity. The shifted coordinate operator ${\bf X}_{i}$ plays a fundamental role in NC
quantum mechanics defined in the $\left(x+\theta\right)$-space, since it is possible to form basis with its eigenvalues.
This possibility is forbidden for the usual coordinate operator ${\bf x}_{i}$ since its components satisfy
nontrivial commutation relations among themselves (\ref{xpcom}). Hence,
differently form ${\bf x}_{i}$, we can say that ${\bf X}_{i}$ forms a basis in 
Hilbert space.
This fact will be important very soon.

With these definitions it seems interesting to study the generators of the group of rotations $SO(D)$.
It is a fact that the usual orbital angular momentum operator
\begin{eqnarray}\label{lij}
{\mathbf \ell}_{ij}\,=\,{\bf x}_{i}{\bf p}_{j}\,-\,{\bf x}_{j}{\bf p}_{i} \; ,
\end{eqnarray}
does not closes in an algebra due to (\ref{xxcom}), that is
\begin{eqnarray}\label{llcomm}
\left[\ell_{ij},\ell_{kl}\right]=i\delta_{il}\ell_{kj}-i\delta_{jl}\ell_{ki}
-i\delta_{ik}\ell_{lj}+i\delta_{jk}\ell_{li}
+i\theta_{il}{\bf p}_{k}{\bf p}_{j}-i\theta_{jl}{\bf p}_{k}{\bf p}_{i}
-i\theta_{ik}{\bf p}_{l}{\bf p}_{j}+i\theta_{jk}{\bf p}_{l}{\bf p}_{i} \; ,
\end{eqnarray}
and so their components cannot be $SO(D)$ generators in this extended Hilbert space.
It is easy to see that the operator
\begin{eqnarray}\label{LXij}
{\bf L}_{ij}\,=\,{\bf X}_{i}{\bf p}_{j}\,-\,{\bf X}_{j}{\bf p}_{i} \; ,
\end{eqnarray}
closes in the $SO(D)$ algebra. Besides, this result can be generalized to the
total angular momentum operator
\begin{eqnarray}\label{Jij}
{\bf J}_{ij}=\,{\bf X}_{i}{\bf p}_{j}\,-\,{\bf X}_{j}{\bf p}_{i}
+\theta_{jl}\pi^{l}_{\;i}-\theta_{il}\pi^{l}_{\;j} \; ,
\end{eqnarray}
that closes the algebra
\begin{eqnarray}\label{JijJklcomm}
\left[{\bf J}_{ij} ,{\bf J}_{kl} \right]=\delta_{il}{\bf J}_{kj}
-\delta_{jl}{\bf J}_{ki}-\delta_{ik}{\bf J}_{lj}+\delta_{jk}{\bf J}_{li} \; ,
\end{eqnarray}
and ${\bf J}_{ij}$ generates rotation in Hilbert space.

Now we return to the discussion about the basis in this NCQM.
It is possible to introduce a continuous basis for a general Hilbert space
searching by a maximal set of commutating operators. The physical coordinates
represented by the positions operators ${\bf x}_{i}$ do not commute and their eigenvalues
cannot be used to form a basis in the Hilbert space ${\cal H}$. This does not occur with
the shifted operators ${\bf X}_{i}$ (\ref{X}), and consequently, their eigenvalues are used
in the construction of such basis. Therefore one can use the shifted position operators
${\bf X}_{i}$ as coordinate basis, although ${\bf x}_{i}$ be the physical position operator.
The noncommutativity of this space stays registered by the presence of the operator
$\theta$ as a spatial coordinate of the system.
A coordinate basis formed by the eigenvectors of $({\bf X},\theta)$ can be introduced, and
for the momentum basis one chooses the eigenvectors of $({\bf p},\pi)$.
Let $|{\bf X}^{\prime},\theta^{\prime} \rangle=|{\bf X}^{\prime}\rangle
\otimes |\theta^{\prime}\rangle$ and $|{\bf p}^{\prime},\pi^{\prime} \rangle$
be the position and momenta states in this $(x+\theta)$-space where the
fundamental relations involving each basis are
\begin{eqnarray}\label{eqautovaloresXtheta}
{\bf X}_{i}|{\bf X}^{\prime},\theta^{\prime} \rangle
= {\bf X}^{\prime}_{i}|{\bf X}^{\prime},\theta^{\prime} \rangle
\hspace{0.2cm} , \hspace{0.9cm} \quad
\theta_{ij}|{\bf X}^{\prime},\theta^{\prime} \rangle
= \theta^{\prime}_{ij}|{\bf X}^{\prime},\theta^{\prime} \rangle
\end{eqnarray}
\begin{eqnarray}\label{eqautovaloresppi}
{\bf p}_{i}|{\bf p}^{\prime},\pi^{\prime} \rangle
= {\bf p}^{\prime}_{i}|{\bf p}^{\prime},\pi^{\prime} \rangle
\hspace{0.2cm} , \hspace{0.9cm} \quad
\pi_{ij}|{\bf p}^{\prime},\pi^{\prime} \rangle
= \pi^{\prime}_{ij}|{\bf p}^{\prime},\pi^{\prime} \rangle
\end{eqnarray}
\begin{eqnarray}\label{ortognalityrelationXtheta}
\langle {\bf X}^{\prime},\theta^{\prime} |{\bf X}^{\prime\prime},\theta^{\prime\prime} \rangle =
\delta^{D}\!\left({\bf X}^{\prime}-{\bf X}^{\prime\prime}\right)\delta^{D(D-1)/2}\!
\left({\bf \theta}^{\prime}-{\bf \theta}^{\prime\prime}\right)
\end{eqnarray}
\begin{eqnarray}\label{ortognalityrelationppi}
\langle {\bf p}^{\prime},\pi^{\prime} |{\bf p}^{\prime\prime},\pi^{\prime\prime} \rangle =
\delta^{D}\!\left({\bf p}^{\prime}-{\bf p}^{\prime\prime}\right)\delta^{D(D-1)/2}\!
\left({\bf \pi}^{\prime}-{\bf \pi}^{\prime\prime}\right)
\end{eqnarray}
\begin{eqnarray}\label{completezarelXtheta}
\int d^{D}{\bf X}^{\prime}d^{D(D-1)/2}\theta^{\prime} |{\bf X}^{\prime},\theta^{\prime} \rangle
\langle {\bf X}^{\prime},\theta^{\prime} | ={\bf 1}
\end{eqnarray}
\begin{eqnarray}\label{completezarelXtheta}
\int d^{D}{\bf p}^{\prime}d^{D(D-1)/2}\pi^{\prime} |{\bf p}^{\prime},\pi^{\prime} \rangle
\langle {\bf p}^{\prime},\pi^{\prime} | ={\bf 1} \; ,
\end{eqnarray}
that are the eigenvalue equation, orthogonality and completeness relations, respectively.
Concerning the last relations and the operators representation we can write that \cite{Amorim1}
\begin{eqnarray}\label{XthetapXtheta}
\langle {\bf X}^{\prime},\theta^{\prime}|{\bf p}_{i}|{\bf X}^{\prime\prime},\theta^{\prime\prime} \rangle =
-i\frac{\partial}{\partial {\bf X}^{\prime\;i}} \delta^{D}\left({\bf X}^{\prime}-{\bf X}^{\prime \prime}\right)
\; \delta^{D(D-1)/2}\left({\bf \theta}^{\prime}-{\bf \theta}^{\prime \prime}\right) \; ,
\end{eqnarray}
and
\begin{eqnarray}\label{XthetapiXtheta}
\langle {\bf X}^{\prime},\theta^{\prime}|\pi_{ij}|{\bf X}^{\prime\prime},\theta^{\prime\prime} \rangle =
-i\delta^{D}\!\left({\bf X}^{\prime}-{\bf X}^{\prime \prime}\right)
\;\frac{\partial}{\partial {\bf \theta}^{\prime\;ij}} \delta^{D(D-1)/2}\left({\bf \theta}^{\prime}
-{\bf \theta}^{\prime \prime}\right) \; .
\end{eqnarray}
It is important to pay attention to the notation.  It is obvious that the prime and double prime notation indicates two different points in $(x+\theta)$-space.  In a few moments, the path integral formalism will need a change in the notation.  But the meaning is the same, i.e., two different points in $(x+\theta)$-space.

The transformations (\ref{XthetapiXtheta}) from one basis to the other are constructed using extended Fourier transforms.
The wave plane defined in this $(x+\theta)$-space is obtained by internal product between position
and momentum states
\begin{eqnarray}\label{waveplane}
\langle {\bf X}^{\prime},\theta^{\prime} |{\bf p}^{\prime\prime},\pi^{\prime\prime} \rangle =
\frac{e^{i\left({\bf p}^{\prime\prime}\cdot{\bf X}^{\prime}+\pi^{\prime\prime}\theta^{\prime}
\right)}}{(2\pi)^{D(D+1)/4}} \; ,
\end{eqnarray}
where ${\bf p}^{\prime\prime}\cdot{\bf X}^{\prime}+\pi^{\prime\prime}\theta^{\prime}=
p_{i}^{\prime\prime}X^{i \; \prime}+\pi_{ij}^{\prime \prime}\theta^{ij \; \prime}/2$.
Others properties of this NCQM are explored in more details
in \cite{Amorim1}.

To provide the interested reader a richer approach, let us, briefly, say some words about the extension of a NCQFT of DFRA spacetime.  Here, the spacetime coordinates $x^{\mu}=(t,{\bf x})$ do not commute itself
satisfying the commutation relation (\ref{xmuxnu}). The parameter $\theta^{\mu\nu}$ is promoted
to coordinate of this spacetime, that in $D=4$, we have six independents spatial coordinates
associated to $\theta^{\mu\nu}$. The commutation relation of the DFRA algebra in Eqs.
(\ref{xpcom})-(\ref{xxcom})-(\ref{xthetacomm})-(\ref{thetapicomm})-(\ref{ppicomm}) can be easily extended
to this NC space
\begin{eqnarray}\label{spacetimealgebraDFRA}
\left[{\mathbf x}_{\mu},{\mathbf x}_{\nu}\right] = i\theta^{\mu\nu}
\hspace{0.2cm} , \hspace{0.2cm}\qquad
\left[{\mathbf x}_{\mu},{\mathbf \theta}_{\nu\alpha}\right] = 0
\hspace{0.2cm} , \hspace{0.2cm}\qquad
\left[{\bf \theta}_{\mu\nu},{\mathbf \theta}_{\alpha\beta}\right] = 0 \; ,
\nonumber \\
\left[{\mathbf x}_{\mu},{\mathbf p}_{\nu} \right] = i\eta_{\mu\nu}
\hspace{0.2cm} , \hspace{0.2cm}\qquad
\left[{\mathbf p}_{\mu},{\mathbf p}_{\nu} \right] = 0
\hspace{0.2cm} , \hspace{0.2cm}\qquad
\left[{\mathbf p}_{\mu},{\mathbf \theta}_{\nu\alpha}\right] = 0 \; ,
\nonumber \\
\left[{\mathbf p}_{\mu},{\mathbf \pi}_{\nu\alpha}\right] = 0
\hspace{0.2cm} , \hspace{0.2cm}\qquad
[{\mathbf x}^{\mu},{\mathbf \pi}_{\nu\rho}]=-{i\over 2}\delta_{\rho\sigma}^{\;\;\;\;\mu\nu}{\mathbf p}_{\nu} \; ,
\hspace{0.5cm}
\end{eqnarray}
where $(p_{\mu},\pi_{\mu\nu})$ are the momenta operators associated to coordinates
$(x^{\mu},\theta^{\mu\nu})$, respectively. The $\theta^{\mu\nu}$ coordinates
are constrained by quantum conditions
\begin{eqnarray}\label{condtheta}
\theta_{\mu\nu}\theta^{\mu\nu}=0
\hspace{0.3cm} \mbox{and} \hspace{0.3cm}
\left(\frac{1}{4}\star\theta_{\mu\nu}\theta^{\mu\nu}\right)^{2}=\lambda_{P}^{8} \; ,
\end{eqnarray}
where $\star\theta_{\mu\nu}=\varepsilon_{\mu\nu\rho\sigma}\theta^{\rho\sigma}$
and $\lambda_{P}$ is the Planck length. In analogy to (\ref{Jij}), the generator
of Lorentz group is
\begin{eqnarray}\label{Mmunu}
{\bf M}_{\mu\nu}=\,{\bf X}_{\mu}{\bf p}_{\nu}\,-\,{\bf X}_{\nu}{\bf p}_{\mu}
+\theta_{\nu\rho}\pi^{\rho}_{\;\mu}-\theta_{\mu\rho}\pi^{\rho}_{\;\nu} \; ,
\end{eqnarray}
and from (\ref{spacetimealgebraDFRA}) we can write the generators
for translations as ${\bf P}_{\mu} = - i \partial_{\mu}\,\,$. 
The shifted coordinate operator $X^{\mu}$ has the analogous definition of (\ref{X}), 
and it satisfies the commutation relations 
\begin{eqnarray}\label{Xmupnu}
[X_{\mu},p_{\nu}]=i\eta_{\mu\nu}
\hspace{0.5cm} \mbox{and} \hspace{0.5cm}
[X_{\mu},X_{\nu}]=0 \; .
\end{eqnarray}
With these ingredients it is easy to construct the commutation relations
\begin{eqnarray}\label{algebraDFR}
\left[ {\mathbf P}_\mu , {\mathbf P}_\nu \right] &=& 0
\hspace{0.2cm} , \nonumber \\
\left[ {\mathbf M}_{\mu\nu},{\mathbf P}_{\rho} \right] &=& \,i\,\big(\eta_{\mu\rho}\,{\mathbf P}_\nu
-\eta_{\mu\nu}\,{\mathbf P}_\rho\big) \; ,
\hspace{0.1cm} \nonumber \\
\left[{\bf M}_{\mu\nu} ,{\bf M}_{\rho\sigma} \right] &=& i\left(\eta_{\mu\sigma}{\bf J}_{\rho\nu}
-\eta_{\nu\sigma}{\bf J}_{\rho\mu}-\eta_{\mu\rho}{\bf J}_{\sigma\nu}+\eta_{\nu\rho}{\bf J}_{\sigma\mu}\right) \; ,
\hspace{-0.5cm}\nonumber \\
\end{eqnarray}
it closes the appropriated algebra, and we can say that ${\mathbf P}_\mu$ and ${\mathbf M}_{\mu\nu}$
are the generators for the extended DFR algebra.
Analyzing the Lorentz symmetry in NCQM following the lines above, 
we can introduce an appropriate theory, for instance, given by a scalar action.  
We know, however, that elementary particles are classified according 
to the eigenvalues of the Casimir operators of the inhomogeneous Lorentz group. 
Hence, let us extend this approach to the Poincar\'e group ${\cal P}$. 
Considering the operators presented here, we can in principle consider that
\begin{eqnarray}\label{ElementoG}
{\mathbf G}={1\over2}\omega_{\mu\nu}{\mathbf M}^{\mu\nu}
-a^\mu{\mathbf p}_{\mu}
+{1\over2}b_{\mu\nu}{\mathbf \pi}^{\mu\nu} \; ,
\end{eqnarray}
is the generator of some group  ${\cal P}'$, which has the Poincar\'e group as a subgroup.  
By defining the dynamical transformation of an arbitrary operator $\mathbf{A}$ 
in ${\cal H}$ in such a way that $\delta \mathbf{A}\,=\,i\,[\mathbf{A},\mathbf{G}]$ 
we arrive at the set of transformations,
\begin{eqnarray}\label{transf}
\delta {\mathbf X}^\mu&=&\omega ^\mu_{\,\,\,\,\nu}{\mathbf X}^\nu+a^\mu\nonumber\\
\delta{\mathbf p}_\mu&=&\omega _\mu^{\,\,\,\,\nu}{\mathbf p}_\nu\nonumber\\
\delta{\mathbf \theta}^{\mu\nu}&=&\omega ^\mu_{\,\,\,\,\rho}{\mathbf \theta}^{\rho\nu}
+ \omega ^\nu_{\,\,\,\,\rho}{\mathbf \theta}^{\mu\rho}+b^{\mu\nu}\nonumber\\
\delta{\mathbf \pi}_{\mu\nu}&=&\omega _\mu^{\,\,\,\,\rho}{\mathbf \pi}_{\rho\nu}
+ \omega _\nu^{\,\,\,\,\rho}{\mathbf \pi}_{\mu\rho}\nonumber\\
\delta {\mathbf M}_1^{\mu\nu}&=&\omega ^\mu_{\,\,\,\,\rho}{\mathbf M}_1^{\rho\nu}
+ \omega ^\nu_{\,\,\,\,\rho}{\mathbf M}_1^{\mu\rho}+a^\mu{\mathbf p}^\nu-a^\nu{\mathbf p}^\mu\nonumber\\
\delta {\mathbf M}_2^{\mu\nu}&=&\omega ^\mu_{\,\,\,\,\rho}{\mathbf M}_2^{\rho\nu}
+ \omega ^\nu_{\,\,\,\,\rho}{\mathbf M_2}^{\mu\rho}+b^{\mu\rho}{\mathbf \pi}_\rho^{\,\,\,\,\nu}
+ b^{\nu\rho}{\mathbf \pi}_{\,\,\,\rho}^{\mu}\nonumber\\
\delta {\mathbf x}^\mu&=&\omega ^\mu_{\,\,\,\,\nu}{\mathbf x}^\nu+a^\mu+{1\over2}b^{\mu\nu}{\mathbf p}_{\nu} \; .
\end{eqnarray}

We observe that there is an unexpected term in the last equation of (\ref{transf}). 
This is a consequence of the coordinate operator in (\ref{X}), which is a nonlinear combination 
of operators that act on different Hilbert spaces.

The action of ${\cal P}'$ over the Hilbert space operators is in some sense equal to the action of the
Poincar\'e group with an additional translation operation on the (${\mathbf \theta}^{\mu\nu}$) sector.
Its generators, all of them, close in an algebra under commutation. Hence, ${\cal P}'$ is a well defined group of transformations. 
As a matter of fact, the commutation of two transformations closes in the algebra
\begin{equation}\label{algebray}
[\delta_2,\delta_1]\,{\mathbf y}=\delta_3\,{\mathbf y} \; ,
\end{equation}
where ${\mathbf y}$ represents any one of the operators appearing in (\ref{transf}). 
The parameters composition rule is given by
\begin{eqnarray}\label{DecompOmega}
\omega^\mu_{3\,\,\nu}&=&\omega^\mu_{1\,\,\,\,\alpha}\omega^\alpha_{2\,\,\,\,\nu}-\omega^\mu_{2\,\,\,\,\alpha}\omega^\alpha_{1\,\,\,\,\nu}\nonumber\\
a_3^\mu&=&\omega^\mu_{1\,\,\,\nu}a_2^\nu-\omega^\mu_{2\,\,\,\nu}a_1^\nu  \\
b_3^{\mu\nu}&=&\omega^\mu_{1\,\,\,\rho}b_2^{\rho\nu}-\omega^\mu_{2\,\,\,\rho}b_1^{\rho\nu}-\omega^\nu_{1\,\,\,\rho}b_2^{\rho\mu}+
\omega^\nu_{2\,\,\,\rho}b_{1}^{\rho\mu} \,\,. \nonumber
\end{eqnarray}
To sum up, the framework showed above demonstrated that in NCQM, the physical coordinates 
do not commute and the respective eigenvectors cannot be used to form a basis 
in ${\cal H}={\cal H}_1 \oplus {\cal H}_2$ \cite{Amorim4}.  This can be accomplished 
with the Bopp shift defined in 
(\ref{X}) with (\ref{Xmupnu}) as consequence.  
So, we can introduce a coordinate basis 
$|X^{\prime}, \theta^{\prime} \rangle = |X^{\prime}\rangle \otimes |\theta^{\prime}\rangle$ in such a way that 
$$X^{\mu}|X^{\prime}, \theta^{\prime}\rangle=X^{\prime\mu}|X^{\prime}, \theta^{\prime}\rangle 
\qquad \mbox{and} \qquad \theta^{\mu\nu}|X^{\prime}, 
\theta^{\prime}\rangle=\theta^{\prime\mu\nu}|X^{\prime},\theta^{\prime}\rangle$$ 
where both satisfies the usual orthonormality and completeness relations.   
In this basis
\begin{eqnarray}\label{XthetapmuXtheta}
\langle X^{\prime},\theta^{\prime}|p_{\mu}| X^{\prime\prime},\theta^{\prime\prime} \rangle \,=
\,-\,i\,\frac{\p}{\p X^{\prime\mu}}\,\delta^{(4)}(X^{\prime}-X^{\prime\prime})
\,\delta^{(6)}(\theta^{\prime} -\theta^{\prime\prime})
\end{eqnarray}
and
\begin{eqnarray}\label{XthetapimunuXtheta}
\langle X^{\prime},\theta^{\prime}|\pi_{\mu}| X^{\prime\prime},\theta^{\prime\prime}\rangle \,
=\,-\,i\,\delta^{(4)}(X^{\prime}-X^{\prime\prime})
\,\frac{\p}{\p \theta^{\prime\mu\nu}}\,\delta^{(6)}(\theta^{\prime} -\theta^{\prime\prime})
\end{eqnarray}
The wave function $\phi(X^{\prime}, \theta^{\prime})=\langle X^{\prime}, \theta^{\prime}|\phi\rangle$ 
represents the physical state $|\phi\rangle$ in the coordinate basis defined above.   
This wave function satisfies some wave equation that can be derived from an action, 
through a variational principle, as usual.

In \cite{Amorim4}, the author constructed directly an ordinary relativistic free quantum theory.  
It was assumed that the physical states are annihilated by the mass-shell condition
\begin{eqnarray}\label{reldisp}
\left(p_{\mu}p^{\mu}-m^2\right)|\phi \rangle = 0 \; ,
\end{eqnarray}
demonstrated through the Casimir operator $C_{1}=p_{\mu}p^{\mu}$ (for more algebraic details see \cite{Amorim4}).  
It is easy to see that in the coordinate representation, this originates the NC KG equation.  
Condition (\ref{reldisp}) selects the physical states that must be invariant under gauge transformations.  
To treat the NC case, let us assume that the second mass-shell condition
\begin{eqnarray}\label{reldisppi}
\left( \pi_{\mu\nu}\pi^{\mu\nu}-\Delta^2 \right)|\phi\rangle = 0
\end{eqnarray}
and must be imposed on the physical states, where $\Delta$ is some constant with dimension $M^4$, 
which sign and value can be defined if $\pi$ is spacelike, timelike or null.  
Analogously the Casimir invariant is $C_{2}=\pi_{\mu\nu}\pi^{\mu\nu}$, demonstrated the validity of (\ref{reldisp}) (see \cite{Amorim4} for details).

Both equations (\ref{reldisp}) and ({\ref{reldisppi}) permit us to construct a general 
expression for the plane wave solution such as \cite{Amorim4}
\begin{eqnarray}\label{phiXtheta}
\phi(X^{\prime},\theta^{\prime})\,\equiv \langle X^{\prime},\theta^{\prime}|\phi\rangle 
\sim \exp \left( ip_{\mu}X^{\prime\mu}+\frac i2 K_{\mu\nu} \theta^{\mu\nu} \right)
\end{eqnarray}
where $p^2\,-\,m^2=0$ and $K^2\,-\,\Delta^2=0$.

In coordinate representation, both (\ref{reldisp}) and (\ref{reldisppi}) are just the 
Klein-Gordon equations, respectively,
\begin{eqnarray}\label{EqKleinGordonp}
\left( \Box_{X}\,-\,m^2 \right) \phi (X^{\prime}, \theta^{\prime} )\,=\,0
\end{eqnarray}
and
\begin{eqnarray}\label{EqKleinGordonpi}
(\Box_{\theta}\,-\,\Delta^2 ) \, \phi (X^{\prime}, \theta^{\prime})\,=\,0
\end{eqnarray}
where $\Box_{X}=\p^{\mu}\,\p_{\mu}$ 
and $\Box_{\theta}=\frac 12\,\p^{\mu\nu}\,\p_{\mu\nu}$, 
$\p_{\mu\nu}=\frac{\p}{\p \theta^{\prime\mu\nu}}$.

Since we constructed the NC KG equation, we will now provide its correspondent action.
An important point is that, due to the noncommutativity of the operator ${\mathbf x}^\mu$ given by Eq. (1) can not
be used to define a possible basis in ${\cal H}$. However, as the components of
${\mathbf X}^\mu$ commute, their
eigenvalues can be used for such purpose, as we said before. 
In \cite{Amorim4} the author has considered these points precisely and have proposed a way for
constructing actions representing possible field theories in this extended
$x+\theta$ spacetime.  One of such actions, generalized in order to allow the
scalar fields to be complex, is given by
\begin{eqnarray}\label{actionscalar}
S(\phi)=\int d^{4}\,X\,d^{6}\theta \, \Big\{\frac{1}{2} \left(\partial_{\mu}\phi\right)^{2} +
{{\lambda^2}\over8}\left(\partial_{\mu\nu}\phi\right)^{2}
-\frac{1}{2}m^2\phi^{2}\Big\} \; ,
\end{eqnarray}

\noindent where $\lambda$ is a parameter with dimension of length, as the Planck
length, which is introduced due to dimensional reasons. 
Besides, it is easy to see that, promoting the limit $\lambda \rightarrow 0$ in Eq. (\ref{actionscalar})
causes the recovering of commutativity \cite{EMCAbreuMJNeves2011}.
Also $\Box= \partial^\mu\partial_\mu  $, with
$\partial_\mu={{\partial}\over{\partial {x}^\mu}}$ and
$\Box_{\theta}={1\over2}\partial^{\mu\nu}\partial_{\mu\nu}$,
where $\partial_{\mu\nu}={{\partial\,\,\,}\over{\partial {\theta}^{\mu\nu}}}\,\,$ and
$\eta^{\mu\nu}=\mbox{diag}(1,-1,-1,-1)$. The corresponding Euler-Lagrange equation reads
\begin{eqnarray}\label{NCKG}
\left(\Box_{X} +\lambda^2\Box_\theta+m^2\right)\phi(X,\theta)=0 \; .
\end{eqnarray}
where $\lambda \rightarrow 0$ recovers commutativity.

Here we revisited the underlying basics to construct a PI
formalism in this NC framework.  It will be present in the section.

\section{The path integral formalism in NC quantum mechanics}
\renewcommand{\theequation}{3.\arabic{equation}}
\setcounter{equation}{0}

To apply the basics of DFRA NCQM  to a path integral formalism,
for simplicity we consider a space of $D=3$, i.e., we have three independent coordinates
associated to $\theta_{ij}$ plus three usual position coordinates ${\bf x}_{i} (i=1,2,3)$.
Thus we will work in a six dimensional space.

The time evolution of a quantum state $|\psi\rangle$ is governed
by the dynamical equation
\begin{eqnarray}\label{eqpsi}
i\frac{d}{dt}|\psi(t)\rangle=H|\psi(t)\rangle \; ,
\end{eqnarray}
where H is the Hamiltonian operator. The solution of equation (\ref{eqpsi}) is the
expression
\begin{eqnarray}\label{solutionPsit}
|\psi(t_{b})\rangle = e^{-iH(t_{b}-t_{a})} |\psi(t_{a})\rangle  \ ,
\end{eqnarray}
that represents the time evolution of a state $|\psi\rangle$ in a time interval $t_{b}-t_{a}\!\!>\!\!0$ between two points $({\bf X}_{a},\theta_{a})$ and $({\bf X}_{b},\theta_{b})$
in this $(x+\theta)-$space, and the Hamiltonian operator $H$ is considered time independent.  As we said before, notice that the indices in $\theta$ like $\theta_a$ and $\theta_b$, for instance, indicate one point in $\theta$-space formed by $(\theta_{12},\theta_{23},\theta_{31})$ coordinates.  It is like the prime and double prime in Eqs. (2.17), (2.18) and (2.21) and so on, which indicates that we are considering two different points in space.  The same notation will be used for $\pi_{\mu\nu}$ in Eq. (3.12) below.

Let $|{\bf X}_{b},\theta_{b}\rangle$ be a position quantum state in $({\bf X}_{b},\theta_{b})$, we operate it in
(\ref{solutionPsit}) to obtain
\begin{eqnarray}\label{psiXbThetab}
\langle {\bf X}_{b},\theta_{b}|\psi(t_{b})\rangle=\langle {\bf X}_{b},\theta_{b}|e^{-iH(t_{b}-t_{a})}|\psi(t_{a})\rangle \; ,
\end{eqnarray}
and introducing the identity
\begin{eqnarray}\label{identityXathetaa}
{\bf 1}=\int d^{3}{\bf X}_{a} d^{3}\theta_{a} \; |{\bf X}_{a},\theta_{a}\rangle \langle {\bf X}_{a},\theta_{a}| \; ,
\end{eqnarray}
we have that
\begin{eqnarray}\label{psiXbthetabGpsiXathetaa}
\langle {\bf X}_{b},\theta_{b}|\psi(t_{b})\rangle=\!\!\int d^{3}{\bf X}_{a} d^{3}\theta_{a} \; \langle {\bf X}_{b},\theta_{b}|e^{-iH(t_{b}-t_{a})}|{\bf X}_{a},\theta_{a}\rangle
\; \langle {\bf X}_{a},\theta_{a}|\psi(t_{a})\rangle \; , \; \; \;
\end{eqnarray}
which can be rewritten as
\begin{eqnarray}\label{psiXbthetabGpsiXathetaa}
\psi(t_{b};{\bf X}_{b},\theta_{b})=\!\!\int d^{3}{\bf X}_{a} d^{3}\theta_{a} \; G(t_{b},{\bf X}_{b},\theta_{b};t_{a},{\bf X}_{a},\theta_{a})
\;
\psi(t_{a};{\bf X}_{a},\theta_{a}) \; . \; \; \;
\end{eqnarray}
This last expression provides the transition of the particle wave-function $\psi$ between the points $({\bf X}_{a},\theta_{a})$ and $({\bf X}_{b},\theta_{b})$ in the $(x+\theta)$ space, and $G$
is the Green-function of this transition
\begin{eqnarray}\label{GXaXb}
G(t_{b},{\bf X}_{b},\theta_{b};t_{a},{\bf X}_{a},\theta_{a}):=\langle {\bf X}_{b},\theta_{b}|e^{-iH(t_{b}-t_{a})}|{\bf X}_{a},\theta_{a}\rangle \; . \;\;\;
\end{eqnarray}
Automatically, the propagator (\ref{GXaXb}) must satisfy the Green equation
\begin{eqnarray}\label{EqGreenG}
\left\{i\frac{\partial}{\partial \tau}-H\left(X,-i\partial_{X};\theta,-i\partial_{\theta}\right)\right\}
G\left(X,\theta;X^{\prime},\theta^{\prime};t \right)=
\delta^{D}(X-X^{\prime})\delta^{D(D-1)/2}(\theta-\theta^{\prime})\delta(\tau) \;\;, \nonumber \\
\end{eqnarray}
in which the Hamiltonian operator $H$ is written in terms of position and momenta operators
discussed in the early section.

The representation as a path integral comes from the sum of all possible transition amplitudes between
points $({\bf X}_{a},\theta_{a})$ and $({\bf X}_{b},\theta_{b})$. We divide the interval time $\delta t=t_{b}-t_{a}$
into $(n+1)$ equal parts of length $\tau$
\begin{eqnarray}\label{deltat}
t_{b}&=&(n+1)\tau+t_{a} \; , \;
\nonumber \\
t_{j}&=&j\tau+t_{a} \;\;\; , \;\; (j=1,...,n) \; .
\end{eqnarray}
Hence, we consider a short transition amplitude between $j-1$ and $j$
in the space $x+\theta$ given by the points $({\bf X}_{j-1},\theta_{j-1})$
and $({\bf X}_{j},\theta_{j})$, where $a<j<b$, for a time interval $\tau:=t_{j}-t_{j-1}$.
The transition amplitude in this short transition is
\begin{eqnarray}\label{Gjj+1}
G(t_{j},{\bf X}_{j},\theta_{j};t_{j-1},{\bf X}_{j-1},\theta_{j-1})=
\langle {\bf X}_{j},\theta_{j}|e^{-iH\tau}|{\bf X}_{j-1},\theta_{j-1}\rangle
 \; , 
\end{eqnarray}
we expand the exponential function for $\tau \ll 1$, and this matrix element is
\begin{eqnarray}\label{Gjj+1expand}
\langle {\bf X}_{j},\theta_{j}|e^{-iH\tau}|{\bf X}_{j-1},\theta_{j-1}\rangle =
\langle {\bf X}_{j},\theta_{j}|{\bf X}_{j-1},\theta_{j-1}\rangle
-i\tau
\langle {\bf X}_{j},\theta_{j}|H|{\bf X}_{j-1},\theta_{j-1}\rangle
+{\cal O}\left(\tau^{2}\right) \; , \nonumber \\
\end{eqnarray}
where we have assumed that ${\bf X}_{0}={\bf X}_{a}$, $\theta_{0}=\theta_{a}$, ${\bf X}_{n+1}={\bf X}_{b}$, $\theta_{n+1}=\theta_{b}$,
$t_{0}=t_{a}$ and $t_{n+1}=t_{b}$ are defined. The representation of $\langle {\bf X}_{j},\theta_{j}|{\bf X}_{j-1},\theta_{j-1}\rangle$
in Fourier transform is
\begin{eqnarray}\label{Xj+1thetaj+1Xjthetaj}
\langle {\bf X}_{j},\theta_{j}|{\bf X}_{j-1},\theta_{j-1}\rangle=\!\! \int\frac{d^{3}{\bf p}_{j}}{(2\pi)^{3}}\frac{d^{3}\Pi_{j}}{(2\pi)^{3}}
\, e^{i{\bf p}_{j}({\bf X}_{j}-{\bf X}_{j-1})+i\Pi_{j}(\theta_{j}-\theta_{j-1})} \; ,
\end{eqnarray}
where $({\bf p}_{j},\Pi_{j})$ are the momenta associated to $({\bf X}_{j},\theta_{j})$, respectively.
If we assume that the DFRA-Hamiltonian operator has the form
\begin{eqnarray}\label{H}
H=\frac{{\bf p}^{2}}{2m}+\frac{\Pi^{2}}{2\Lambda}+V({\bf X},\theta)  \ \ ,
\end{eqnarray}
the matrix element $\langle {\bf X}_{j},\theta_{j}|H|{\bf X}_{j-1},\theta_{j-1}\rangle$
is calculated to obtain the expression
\begin{eqnarray}\label{Xj+1Hxj}
\langle {\bf X}_{j},\theta_{j}|H|{\bf X}_{j-1},\theta_{j-1}\rangle =
\int\frac{d^{3}{\bf p}_{j}}{(2\pi)^{3}}\frac{d^{3}\Pi_{j}}{(2\pi)^{3}}
\,H({\bf p}_{j},\Pi_{j},{\bf X}_{j-1},\theta_{j-1})
\;e^{i{\bf p}_{j}({\bf X}_{j}-{\bf X}_{j-1})+i\Pi_{j}(\theta_{j}-\theta_{j-1})} \; , \nonumber \\
\end{eqnarray}
where the particle is subjected to a potential $V$ that depends on the variables $({\bf X},\theta)$,
and now $(p,\Pi,{\bf X},\theta)$ are eigenvalues associated to their operators. Using these results,
the transition amplitude (\ref{Gjj+1}) is given by
\begin{eqnarray}\label{Gjj+1eH}
\langle {\bf X}_{j},\theta_{j}|e^{-iH\tau}|{\bf X}_{j-1},\theta_{j-1}\rangle =
\int \frac{d^{3}{\bf p}_{j}}{(2\pi)^{3}}\frac{d^{3}\Pi_{j}}{(2\pi)^{3}}
\; e^{i\left[{\bf p}_{j}\left({\bf X}_{j}-{\bf X}_{j-1}\right)+\Pi_{j}\left(\theta_{j}-\theta_{j-1}\right)-\tau H\left({\bf p}_{j},{\bf X}_{j-1};\Pi_{j},\theta_{j}\right) \right]}  \; . \nonumber \\
\end{eqnarray}
The Feynman path integral is obtained summing all contributions of paths between
the points $({\bf X}_{a},\theta_{a})$ and $({\bf X}_{b},\theta_{b})$
\begin{eqnarray}\label{SumGXjXj-1}
&&\langle {\bf X}_{b},\theta_{b}|e^{-iH(t_{b}-t_{a})}|{\bf X}_{a},\theta_{a}\rangle \nonumber \\
&=&\lim_{n\rightarrow \infty}\int \prod_{j=1}^{n} d^{3}{\bf X}_{j} d^{3}\theta_{j}
\int \prod_{j=1}^{n+1} \frac{d^{3}{\bf p}_{j}}{(2\pi)^{3}}\frac{d^{3}\Pi_{j}}{(2\pi)^{3}}
\exp\!\left(\!i\!\sum_{j=1}^{n+1}\left[{\bf p}_{j}\left({\bf X}_{j}-{\bf X}_{j-1}\right)
\right. \right. \nonumber \\
&+&\left. \left.
\Pi_{j}\left(\theta_{j}-\theta_{j-1}\right)
-\left(t_{j}-t_{j-1}\right) H\left({\bf p}_{j},{\bf X}_{j-1};\Pi_{j}\theta_{j-1}
\right)\right]\phantom{\sum_{j=1}^{n+1}} \hspace{-0.6cm}\right)
\end{eqnarray}
where the limit $n\rightarrow\infty$ has been taken.   This result can be written in the compact form
\begin{eqnarray}\label{Gintfuncional}
\langle {\bf X}_{b},\theta_{b}|e^{-iH(t_{b}-t_{a})}|{\bf X}_{a},\theta_{a}\rangle=
\!\int {\cal D}{\bf X} {\cal D}{\bf p}
{\cal D}\theta {\cal D}\Pi
\,\exp \left[i\int_{t_{a}}^{t_{b}}\!dt\left({\bf p}\cdot\dot{{\bf X}}+\Pi\dot{\theta}-H({\bf X},{\bf p};\theta,\Pi)\right)\right] \; , \nonumber \\
\end{eqnarray}
where $H$ is DFRA-Hamiltonian of the system
\begin{eqnarray}\label{Hautovalores}
H({\bf X},{\bf p};\theta,\Pi)=\frac{{\bf p}^{2}}{2m}+\frac{\Pi^{2}}{2\Lambda}+V({\bf X},\theta) \; ,
\end{eqnarray}
and now $H$ is only written in terms of the eigenvalues.
The integrals (\ref{Gintfuncional}) are functional integrations over the entire phase space
of the configuration, with the boundary conditions
${\bf X}(t=t_{a})={\bf X}_{a}$, $\theta(t=t_{a})=\theta_{a}$, ${\bf X}(t=t_{b})={\bf X}_{b}$ and $\theta(t=t_{b})=\theta_{b}$.
If the Hamiltonian is of the form (\ref{H}), it is convenient to perform the momentum integrations $(p,\Pi)$
in (\ref{SumGXjXj-1}). Shifting the integration variables ${\bf p}_{j} \rightarrow {\bf p}_{j}-m\Delta {\bf X}_{j}/\tau$
and $\Pi_{j} \rightarrow \Pi_{j}-m\Delta \theta_{j}/\tau$, the results of momentum integrals
substituted in (\ref{SumGXjXj-1}) give us
\begin{eqnarray}\label{intFuncDxDtheta}
\langle {\bf X}_{b},\theta_{b}|e^{-iH(t_{b}-t_{a})}|{\bf X}_{a},\theta_{a}\rangle=
N\int \frac{{\cal D}{\bf X}}{2\pi} \frac{{\cal D}\theta}{(2\pi)^{3}}e^{iS\left({\bf X},\theta\right)}
 , \;\;\;\;
\end{eqnarray}
that it has the form of a functional integral over configuration space,
in which $S\left({\bf X},\theta\right)$ is the action integral of the system
\begin{eqnarray}\label{action}
S\left({\bf X},\theta\right)=\int_{t_{a}}^{t_{b}}dt \; L(\dot{{\bf X}},\dot{\theta}) \; ,
\end{eqnarray}
and this integrating term is the Lagrangian function of the system
\begin{eqnarray}\label{Lagrangianfree}
L({\bf X},\dot{{\bf X}};\theta,\dot{\theta})=\frac{1}{2}m\dot{{\bf X}}^{2}
+\frac{1}{2}\Lambda\dot{\theta}^{2}-V\left({\bf X},\theta\right) \; .
\end{eqnarray}
The factor $N$ is just a normalization constant. As we expected,
the representation of the path integral is given by functional
integration over the configuration space-$(x+\theta)$ of the
exponential function of the action integral.
Naturally, what emerges in this result is the DFRA-Lagrangian function of the system.
In the next section we apply it to some simple examples, as free particle and the 
isotropic harmonic oscillator.

\section{Examples}
\renewcommand{\theequation}{4.\arabic{equation}}
\setcounter{equation}{0}

In this section we will exemplify the formalism, developed before, using two simple systems: the NC free particle and the NC harmonic oscillator.
For the first case the Lagrangian of the free particle is
\begin{eqnarray}\label{Hlivre}
L(\dot{{\bf X}};\dot{\theta})=\frac{1}{2}m\dot{{\bf X}}^{2}
+\frac{1}{2}\Lambda\dot{\theta}^{2} \; ,
\end{eqnarray}
where $\Lambda$ is a dimensionfull parameter \cite{Amorim1} and we can write the free propagator in the product representation
\begin{eqnarray}\label{GFreeProd}
K_{0}({\bf X}_{b},\theta_{b};{\bf X}_{a},\theta_{a},\tau)&=&
\lim_{n\rightarrow \infty}\left(\frac{m}{i\tau}\right)^{3(n+1)/2} 
\;\int \prod_{j=1}^{n}d^{3}{\bf X}_{j}
\exp\left[\frac{im}{2\tau}\sum_{j=1}^{n+1}\left({\bf X}_{j}-{\bf X}_{j-1}\right)^{2}\right]
\hspace{0.5cm}\nonumber \\
&\times& \left(\frac{\Lambda}{i\tau}\right)^{\!\!3(n+1)/2} \!\!\!\! \int \prod_{j=1}^{n}d^{3}\theta_{j} \exp\left[\frac{i\Lambda}{2\tau}\sum_{j=1}^{n+1}\left(\theta_{j}-\theta_{j-1}\right)^{2}\right] .
\hspace{0.25cm}
\end{eqnarray}
Using the known result in the literature for the integrals
\begin{eqnarray}\label{identidadeint}
\int \prod_{j=1}^{n}d^{3}{\bf X}_{j}
\exp\left[\frac{im}{2\tau}\sum_{j=1}^{n+1}\left({\bf X}_{j}-{\bf X}_{j-1}\right)^{2}\right]
=\frac{\left(i\tau/m\right)^{3n/2}}{(n+1)^{3/2}}\;\exp\left[\frac{im}{2(n+1)\tau}\left({\bf X}_{b}-{\bf X}_{a}\right)^{2} \right] \; ,
\end{eqnarray}
and the analogous for the $\theta^{\mu\nu}$ part, we obtain the expression of the free propagator
\begin{eqnarray}\label{GFree}
K_{0}(t_{b},{\bf X}_{b},\theta_{b};t_{a},{\bf X}_{a},\theta_{a})&=&i\Theta(t_{b}-t_{a})\left[\frac{m\Lambda}{(t_{b}-t_{a})^{2}}\right]^{3/2} \nonumber \\
&\times& \exp\left[\frac{im\left({\bf X}_{b}-{\bf X}_{a}\right)^{2}+i\Lambda\left(\theta_{b}-\theta_{a}\right)^{2}}{2(t_{b}-t_{a})} \right] , \hspace{0.7cm} \nonumber \\
\mbox{}
\end{eqnarray}
with the condition $t_{b}-t_{a}>0$ which is satisfied by causality.

A second example we have the Lagrangian of the NC isotropic harmonic oscillator \cite{Amorim1}
\begin{eqnarray}\label{LMHS}
L({\bf X},\dot{{\bf X}};\theta,\dot{\theta})=\frac{1}{2}m\dot{{\bf X}}^{2}
+\frac{1}{2}\Lambda\dot{\theta}^{2}-\frac{1}{2}m\omega^{2}{\bf X}^{2}
-\frac{1}{2}\Lambda\Omega^{2}{\bf \theta}^{2} \; ,
\hspace{-0.6cm}
\end{eqnarray}
where $\omega$ and $\Omega$ are the oscillation frequencies in the spaces $({\bf X},\theta)$, respectively.
The discrete form of the path integral is
\begin{eqnarray}\label{GOHS}
& & K({\bf X}_{b},\theta_{b};{\bf X}_{a},\theta_{a},\tau) \nonumber \\
&=&\lim_{n\rightarrow \infty}\left(\frac{m}{i\tau}\right)^{3(n+1)/2}
\!\!\!\int \prod_{j=1}^{n}d^{3}{\bf X}_{j}
\prod_{j=1}^{n+1}
\exp\left\{i\tau\left[\frac{m}{2}\left(\frac{{\bf X}_{j}-{\bf X}_{j-1}}{\tau}\right)^{2}\!\!\!\!-\frac{1}{2}m\omega^{2}\left(\frac{{\bf X}_{j}+{\bf X}_{j-1}}{2}\right)^{\!2}\right]\right\}
\nonumber \\
&\times& \lim_{n\rightarrow \infty} \left(\frac{\Lambda}{i\tau}\right)^{3(n+1)/2} \!\!\! \int \prod_{j=1}^{n}d^{3}\theta_{j}\prod_{j=1}^{n+1}
\!\!\times\exp\left\{i\tau\left[\frac{\Lambda}{2}\left(\frac{{\bf \theta}_{j}-{\bf \theta}_{j-1}}{\tau}\right)^{2}\!\!\!\!-\frac{1}{2}\Lambda\Omega^{2}\left(\frac{{\bf \theta}_{j}+{\bf \theta}_{j-1}}{2}\right)^{\!2}\right]  \right\} \; , \nonumber \\
\end{eqnarray}
and using the same method for the usual commutative model we obtain that
\begin{eqnarray}\label{propagadorOHS}
&&K({\bf X}_{b},\theta_{b};{\bf X}_{a},\theta_{a},\tau) \nonumber \\
&=&i\left(\frac{m\omega\Lambda\Omega}{4\pi^{2} \sin(\omega\tau) \sin(\Omega\tau)}\right)^{3/2} 
\exp\left[\frac{im\omega}{2\sin(\omega\tau)} \left(\cos(\omega\tau)\left({\bf X}_{b}^{2}+{\bf X}_{a}^{2} \right)-2{\bf X}_{a}\cdot{\bf X}_{b} \right)\right]
\nonumber \\
&\times&\exp\left[\frac{i\Lambda\Omega}{2\sin(\Omega\tau)} \left(\cos(\Omega\tau)\left({\bf \theta}_{b}^{2}+{\bf \theta}_{a}^{2} \right)-2{\bf \theta}_{a}\cdot{\bf \theta}_{b} \right)\right] \; , \;\;\;
\end{eqnarray}
where $\tau$ is defined as $\tau:=t_{b}-t_{a}$.


The partition function of the NC isotropic harmonic oscillator can be obtained by the usual definition
\begin{eqnarray}\label{PartitionZ}
Z(\beta):={\mbox Tr}e^{-\beta H} \; ,
\end{eqnarray}
in which the trace operation is taken over the continuous set of eigenstates
of the position operators $|{\bf X},\theta \rangle$
\begin{eqnarray}\label{Zint}
Z(\beta)=\int d^{3}{\bf X}d^{3}\theta \; \langle {\bf X},\theta | e^{-\beta H} | {\bf X},\theta \rangle \; ,
\end{eqnarray}
which in terms of the propagator we can write that ${\bf X}_{a}={\bf X}_{b}={\bf X}$, $\theta_{a}=\theta_{b}=\theta$
and $\beta=i\tau$ as the imaginary time interval, we have
\begin{eqnarray}\label{Zint}
Z(\beta)=\int d^{3}{\bf X}d^{3}\theta \; K\left({\bf X},\theta;{\bf X},\theta,-i\beta \right) \; .
\end{eqnarray}
Substituting the propagator (\ref{GOHS}), after a trivial Gaussian integration, we obtain that
\begin{eqnarray}\label{ZOHS}
Z(\beta)=\frac{1}{16\sinh^{3}\left(\frac{\omega\beta}{2}\right)\sinh^{3}\left(\frac{\Omega\beta}{2}\right)} \; .
\end{eqnarray}
We can see clearly that the NC space contribution is through the $\Omega$ frequency.
The mean energy deduced from the partition function is given by the formula
\begin{eqnarray}\label{meanE}
\left. \langle E \rangle =-\frac{\partial }{\partial \beta} \log Z(\beta) \right|_{\beta=\frac{1}{T}} \; ,
\end{eqnarray}
where the $\beta$ parameter is identified as the inverse of the temperature $T$, and we have the result
\begin{eqnarray}\label{PlanckE}
\langle E \rangle = \frac{3}{2}\left(\omega+\Omega\right)+\frac{3\omega}{e^{\omega/T}-1}+\frac{3\Omega}{e^{\Omega/T}-1} \; .
\end{eqnarray}
This is the Planck's formula for the average energy of the oscillator. At very low temperatures $(T\ll \omega)$
and $(T\ll \Omega)$, we have the ground state
\begin{eqnarray}\label{EPlanckTpequeno}
\langle E \rangle \approx \frac{3}{2}\left(\omega+\Omega\right) \; ,
\end{eqnarray}
and at high temperature we obtain the classical Boltzmann statistics
\begin{eqnarray}\label{EPlanckTgrande}
\langle E \rangle \approx 6T \; ,
\end{eqnarray}
in which the factor $6$ sets the spatial dimensions in the space-$({\bf x}+\theta)$.
In the next section we apply the path integral (\ref{intFuncDxDtheta}) to the framework
of quantum field theory in the NC $DFRA$ spacetime discussing the generating
functional, perturbation theory and $n$-points Green functions.

\section{The effective action and Green functions}
\renewcommand{\theequation}{5.\arabic{equation}}
\setcounter{equation}{0}

We apply the path integral method to calculate the matrix elements
of the position operators $({\bf X},\theta)$ in DFRA space.
For the ordered product of $n$ operators ${\bf X}$ and $m$ operators
$\theta$ we will introduce the expression 
\begin{eqnarray}\label{elementmatrix}
&&\langle {\bf X}_{b},\theta_{b},t_{b}|T\left[{\bf X}(t_{1})\ldots{\bf X}(t_{n})\theta(t_{1}^{\prime})\ldots\theta(t_{m}^{\prime})\right]|{\bf X}_{a},\theta_{a},t_{a}\rangle \nonumber \\
&&\qquad\qquad\qquad\qquad\qquad\qquad=\;\int \frac{{\cal D}{\bf X}}{2\pi} \frac{{\cal D}{\bf \theta}}{(2\pi)^{3}} {\bf X}(t_{1})\ldots{\bf X}(t_{n})\theta(t_{1}^{\prime})\ldots\theta(t_{m}^{\prime})\;e^{iS({\bf X},\theta)} \; , \nonumber \\
\end{eqnarray}
and in the case of two operators $({\bf X},\theta)$ we have that
\begin{eqnarray}\label{elementmatrixXtheta}
\int \frac{{\cal D}{\bf X}}{2\pi} \frac{{\cal D}{\bf \theta}}{(2\pi)^{3}} {\bf X}(t_{1})\theta(t_{1}^{\prime})\;e^{iS({\bf X},\theta)}
=\left\{
  \begin{array}{l l}
     \langle {\bf X}_{b},\theta_{b},t_{b}|{\bf X}(t_{1})\theta(t_{1}^{\prime})|{\bf X}_{a},\theta_{a},t_{a}\rangle \; , \; t_{1}>t_{1}^{\prime} \\
     \langle {\bf X}_{b},\theta_{b},t_{b}|\theta(t_{1}^{\prime}){\bf X}(t_{1})|{\bf X}_{a},\theta_{a},t_{a}\rangle \; , \; t_{1}<t_{1}^{\prime}   \\
\end{array} \right. \;\; .
\end{eqnarray}
The transition amplitude in the presence of external sources $(J_{i},\eta_{ij})$ can be written as
\begin{eqnarray}\label{amplitudeJeta}
\langle {\bf X}_{b},\theta_{b},t_{b}|{\bf X}_{a},\theta_{a},t_{a}\rangle^{(J,\eta)}
=\!\int \frac{{\cal D}{\bf X}}{2\pi} \frac{{\cal D}{\bf \theta}}{(2\pi)^{3}}
\exp\left\{i\!\int_{{t}_{a}}^{t_{b}} \!\! dt \left[L+{\bf J}(t)\cdot{\bf X}(t)+{\bf \eta}(t)\cdot\theta(t) \right] \right\} \; ,
\hspace{-1cm}\nonumber \\
\end{eqnarray}
where $\eta_{ij}$ is an antisymmetric external source associated to coordinate $\theta_{ij}$, and $J$.
The transition amplitude (\ref{amplitudeJeta}) can be used as a generating functional of the matrix elements of the position operators, that are given
by its functional derivatives with respect to $(J,\eta)$
\begin{eqnarray}\label{derivativesJ}
&&\langle {\bf X}_{b},\theta_{b},t_{b}|T\left[{\bf X}(t_{1})...{\bf X}(t_{n})\theta(t_{1}^{\prime})\ldots\theta(t_{m}^{\prime})\right]|{\bf X}_{a},\theta_{a},t_{a}\rangle
\nonumber \\
&&\left.=\left(\frac{1}{i}\right)^{n+m}\!\!\!\!\!\!\!\!\!\frac{\delta^{n+m}
}{\delta J(t_{1})\ldots\delta J(t_{n})\delta \eta(t_{1}^{\prime})\ldots\delta \eta(t_{m}^{\prime})}
\times \langle {\bf X}_{b},\theta_{b},t_{b}|{\bf X}_{a},\theta_{a},t_{a}\rangle^{(J,\eta)}\right|_{(J,\eta)=0}\; .
\hspace{0.7cm}
\end{eqnarray}
We can relate the correlation function (\ref{elementmatrixXtheta}) for two operators
to the definition of the vacuum expectation value by
\begin{eqnarray}\label{vacuumexp}
&&\langle 0|T\left[{\bf X}(t_{1})...{\bf X}(t_{n})...\theta(t_{1}^{\prime})...\theta(t_{m}^{\prime})\right]|0\rangle=
\nonumber \\
&&\lim_{t_{a}\rightarrow-i\infty \atop t_{b}\rightarrow +i\infty}
\frac{\int {\cal D}{\bf X} {\cal D}{\theta} {\bf X}(t_{1})...{\bf X}(t_{n})...\theta(t_{1}^{\prime})...\theta(t_{m}^{\prime}) e^{i\int_{t_{a}}^{t_{b}} dt \; \left[L+i\varepsilon {\bf X}^{2}(t)+i\varepsilon \theta^{2}(t)\right] } }{\int {\cal D}{\bf X} {\cal D}{\theta} e^{i\int_{t_{a}}^{t_{b}} dt \; \left[L+i\varepsilon {\bf X}^{2}(t)+i\varepsilon \theta^{2}(t) \right] } } \; ,
\hspace{-1.2cm}\nonumber \\
\end{eqnarray}
where we have performed a Wick rotation in the temporal axis. Two extra terms $i\varepsilon {\bf X}^{2}$ and
$i\varepsilon {\bf \theta}^{2}$ have been added to guarantee the convergence of the path integral. Those extra
terms with the sign $(+)$ are underlying to generate the causal Green functions (Feynman's Green functions).
The last formula permits us to define an important ingredient to generate the Green functions of the theory. We define
the generating functional of the Green functions as
\begin{eqnarray}\label{ZJeta}
Z(J,\eta)\!:=\!\frac{\int {\cal D}{\bf X} \; {\cal D}{\theta} \; e^{i\int_{-\infty}^{+\infty} dt
\left[L+i\varepsilon {\bf X}^{2}(t)+i\varepsilon \theta^{2}(t)+{\bf J}(t)\cdot {\bf X}(t)
+{\bf \eta}(t)\cdot{\bf \theta}(t)\right]}}{\int {\cal D}{\bf X} {\cal D}{\theta} \;
e^{i\int_{-\infty}^{+\infty} dt \left[L+i\varepsilon {\bf X}^{2}(t)+i\varepsilon \theta^{2}(t) \right] } } \; ,
\hspace{-1.1cm}\nonumber \\
\end{eqnarray}
where external sources $({\bf J},\eta)$ are introduced to calculate the $(x+\theta)$-space $n$-points Green functions. These expression is related to the vacuum expectation value by the operation concerning the sources derivation
\begin{eqnarray}\label{correlationZ}
\langle 0|T\left[{\bf X}(t_{1})...{\bf X}(t_{n})...\theta(t_{1}^{\prime})...\theta(t_{m}^{\prime})\right]|0\rangle 
&=&\left(\frac{1}{i}\right)^{n+m}\!\!\!\!\!\!\!\!\!\frac{\delta^{n+m}Z({\bf J},\eta)
}{\delta J(t_{1})...\delta J(t_{n}) ... \delta \eta(t_{1}^{\prime})...\delta \eta(t_{m}^{\prime})} \\
&=&\left.\!\frac{\int {\cal D}{\bf X} \; {\cal D}{\theta} {\bf X}(t_{1})...{\bf X}(t_{n})
...\theta(t_{1}^{\prime})...\theta(t_{m}^{\prime}) e^{iS({\bf X},\theta) } }
{\int {\cal D}{\bf X} {\cal D}{\theta} \; e^{iS({\bf X},\theta)}}\right|_{J,\eta=0} \; .
\nonumber 
\end{eqnarray}
We pass all those definitions for the approach of quantum field theory. The generating functional
for a free massive scalar field is given by
\begin{eqnarray}\label{ZJescalar}
Z_{0}(J)=\frac{\int {\cal D}\phi \; e^{i\int d^{4}X\,d^{6}\theta\,W(\theta)
\left[{\mathcal L}_{0}+\frac{i\varepsilon}{2} \phi^{2}+J({\bf X},\theta)\phi(X,\theta)\right]}}{\int {\cal D}\phi \;
e^{i\int d^{4}X\,d^{6}\theta\,W(\theta) \left({\mathcal L}_{0}+\frac{i\varepsilon}{2} \phi^{2} \right) } } \; ,
\end{eqnarray}
where the weighting $W(\theta)$ (positive) function will permits us to work with series expansions in $\theta$, i.e., with truncated power series expansion of functions of $\theta$.  For any large $\theta_{\mu\nu}$ it falls to zero quickly so that all integrals are well defined.  For details see \cite{Carlson,Morita}.
The functional is normalized $Z(J=0)=1$, and
${\mathcal L}_{0}$ is the Lagrangian with no interaction
of the massive scalar field
\begin{eqnarray}\label{Lescalar}
{\cal L}_{0}(\phi,\partial_{\mu}\phi,\partial_{\mu\nu}\phi)=\frac{1}{2}\partial_{\mu}\phi\partial^{\mu}\phi
+\frac{\lambda^{2}}{8}\partial_{\mu\nu}\phi\partial^{\mu\nu}\phi-\frac{1}{2}m^{2}\phi^{2} \;.
\nonumber \\
\end{eqnarray}
in the $(X+\theta)$-space. The action of the scalar field has been modified in
the generating functional by the terms of source and $+i\varepsilon\phi^{2}/2$,
the variational principle give us the field equation
\begin{eqnarray}\label{KleinGordonJ}
\left(\Box+\lambda^{2}\Box_{\theta}+m^{2}-i\varepsilon\right)\phi(X,\theta)=J(X,\theta) \; ,
\end{eqnarray}
where $\Box_{\theta}=\partial_{\mu\nu}\partial^{\mu\nu}/4$. The solution for this equation
in the presence of an external source is due to the Green method
\begin{eqnarray}\label{solutionphi}
\phi(X,\theta)=\phi_{0}(X,\theta)
+\!\!\int d^{4}xd^{6}\theta \,W(\theta)\; G(X,\theta;X^{\prime},\theta^{\prime})J(X^{\prime},\theta^{\prime}) \; ,
\end{eqnarray}
in which $\phi_{0}$ is the solution for the field equation with no source and $G$ is the causal Green
function (Feynman) that satisfies the following equation
\begin{eqnarray}\label{EqGreen}
\left(\Box+\lambda^{2}\Box_{\theta}+m^{2}-i\varepsilon\right)G(X,\theta,X^{\prime},\theta^{\prime})
=\delta^{(4)}(X-X^{\prime})\delta^{(6)}(\theta-\theta^{\prime}) \; .
\end{eqnarray}
The Green function can be calculated as the inverse of the Klein-Gordon operator
\begin{eqnarray}\label{XthetaGXtheta}
G(X,\theta;X^{\prime},\theta^{\prime})
=\langle X,\theta  | \frac{1}{\Box+\lambda^{2}\Box_{\theta}+m^{2}-i\varepsilon}
| X^{\prime},\theta^{\prime}\rangle \; , \;.
\end{eqnarray}
Including the completeness relation (\ref{completezarelXtheta}) and identifying
the differential representation of the momenta operators $\partial_{\mu} \rightarrow iP_{\mu}$
and $\partial_{\mu\nu} \rightarrow iK_{\mu\nu}$, we obtain the
Fourier representation
\begin{eqnarray}\label{GFourier}
G(X,\theta;X^{\prime},\theta^{\prime})=\int \frac{d^{4}p}{(2\pi)^{4}}\frac{d^{6}K}{(2\pi)^{6}}
\frac{e^{ip.(X-X^{\prime})+iK.(\theta-\theta^{\prime})}}{p^{2}+\lambda^{2}K^{2}-m^{2}+i\varepsilon} \; ,
\end{eqnarray}
of the Feynman Green function.

The action of the massive scalar field associated with the Lagrangian (\ref{Lescalar}) with
an external source can be rewritten in the form of field-operator-field by one simple integration by parts
\begin{eqnarray}\label{actionphiJ}
S^{J}(\phi)=-\int d^{4}xd^{6}\theta \left[\frac{1}{2}\phi\left(\Box
+\lambda^{2}\Box_{\theta}+m^{2}-i\varepsilon\right)\phi
-J\phi\right] \; ,
\end{eqnarray}
and substituting the solution (\ref{solutionphi}) with help of the Green equation (\ref{EqGreen})
we obtain
\begin{eqnarray}\label{SJSGreen}
S^{J}(\phi)=S(\phi)+\frac{1}{2}\int d^{4}xd^{6}\theta
d^{4}x^{\prime}d^{6}\theta^{\prime} \,W(\theta)\,W(\theta^{\prime})\,J(X,\theta)G(X,\theta;X^{\prime},\theta^{\prime})
J(X^{\prime},\theta^{\prime}) \; .
\end{eqnarray}
This expression can be substituted into the generating functional.  So, we can write it in terms
of the source and the causal Green function
\begin{eqnarray}\label{ZJGJ}
Z_{0}(J)=\exp\left[\frac{i}{2}\int d^{4}xd^{6}\theta
d^{4}x^{\prime}d^{6}\theta^{\prime}
 J(X,\theta)G(X,\theta;X^{\prime},\theta^{\prime})\,W(\theta)\,W(\theta^{\prime})\,
J(X^{\prime},\theta^{\prime})  \right] \; .
\end{eqnarray}
The combination of an interaction term ${\cal L}_{int}(\phi)$ into the Lagrangian
(\ref{Lescalar}) lead us to rewrite the generating functional as
\begin{eqnarray}\label{ZJint}
Z(J)=\frac{\int {\cal D}\phi \; e^{i\int d^{4}X\,d^{6}\theta\,W(\theta)\,
\left[{\mathcal L}_{0}+{\cal L}_{int}(\phi)+\frac{i\varepsilon}{2} \phi^{2}+J\phi\right]}}{\int {\cal D}\phi \;
e^{i\int d^{4}X\,d^{6}\theta \,W(\theta)\,\left[{\mathcal L}_{0}+{\cal L}_{int}(\phi)+\frac{i\varepsilon}{2} \phi^{2} \right] } } \; ,
\end{eqnarray}
where we have assumed that the interaction Lagrangian ${\cal L}_{int}(\phi)$ depends on $\phi$,
that is, it is a self-interaction $\phi^{n}$, with $n\geq 3$ integer. It is easy to see that
with this assumption the generating functional can be written as
\begin{eqnarray}\label{ZJintseries}
Z(J)=\frac{\exp \left[i\int d^{4}xd^{6}\theta \,W(\theta)\,{\cal L}_{int}\left(\frac{\delta}{\delta J} \right) \right]Z_{0}(J)}
{\left.\exp \left[i\int d^{4}xd^{6}\theta \,W(\theta)\,{\cal L}_{int}\left(\frac{\delta}{\delta J} \right) \right]Z_{0}(J)\right|_{J=0}}
=
\nonumber \\
=N\exp \left[i\int d^{4}xd^{6}\theta {\cal L}_{int}\left(\frac{\delta}{\delta J} \right) \right]Z_{0}(J) \; ,
\end{eqnarray}
in which $N$ is a normalization factor. This expression represents the perturbation series of a NCQFT.
The generating functional in the presence of an interaction is related to those in the free theory.
The normalization factor $N$ is within the series inside the denominator of (\ref{ZJintseries}) and represents
the vacuum diagrams that are canceled with the contribution of the numerator.

The analogous relation to (\ref{correlationZ}) for a
QFT theory leads us to define the $n$-points Green function
in $(X+\theta)$-space
\begin{eqnarray}\label{Gn}
G^{(n)}((X_{1},\theta_{1}), ... ,(X_{n},\theta_{n}))
\,&=&\,\langle 0|T\left[\Phi(X_{1},\theta_{1})...\Phi(X_{n},\theta_{n})\right]|0\rangle \nonumber \\
&=&\left.\left(\frac{1}{i}\right)^{n}\!\!\frac{\delta^{n}Z(J)}{\delta J(X_{1},\theta_{1})...\delta J(X_{n},\theta_{n})}\right|_{J=0}\;\;.
\end{eqnarray}

%
\noindent In this last equation the correlation function is written in terms of scalar field operators $\Phi$.
Substituting this result in expression (\ref{ZJGJ}), the two-points free Green function is
\begin{eqnarray}\label{Gn=2}
\left.G^{(2)}((X_{1},\theta_{1}),(X_{2},\theta_{2}))=
- \frac{\delta^{2}Z_{0}(J)}{\delta J(X_{1},\theta_{1})\delta J(X_{2},\theta_{2})} 
= G(X_{1},\theta_{1};X_{2},\theta_{2})\right|_{J=0} \; ,
\end{eqnarray}
Consequently, it has the Fourier representation (\ref{GFourier}).
For the interaction picture, the definition (\ref{Gn}) generates the connected
and non connected NC Green functions. Since we wish to construct the one-particle-irreducible
(1PI) NC Green functions, we define the generating functional of the connected Green
functions
\begin{eqnarray}\label{WJ}
W(J)=-\ln Z(J) \; ,
\end{eqnarray}
and the effective action is defined by the Legendre transformation
\begin{eqnarray}\label{Gamma}
\Gamma(\phi)=W(J)-\int d^{4}xd^{6}\theta J(x,\theta)\phi(x,\theta) \; .
\end{eqnarray}
The 1PI NC Green functions are the functional derivatives
of the effective action which relation to classical field $\phi$ is
\begin{eqnarray}\label{Gamman}
\left.\Gamma^{(n)}((X_{1},\theta_{1})...(X_{n},\theta_{n}))=
\frac{\delta^{n} \Gamma(\phi)}{\delta\phi(X_{1},\theta_{1})...\delta\phi(X_{n},\theta_{n})}\right|_{\phi=0}
 .
\hspace{-0.7cm}\nonumber \\
\end{eqnarray}

It is subtle to realize that to obtain the commutative 1PI Green functions we cannot simply use the standard procedure and make the limit $\theta \rightarrow 0$.  This happens, of course, because $\theta^{\mu\nu}$ is not a parameter.  It is a coordinate in $(x+\theta)$-space.  To obtain commutativity we have to make the parameter $\lambda \rightarrow 0$ in Eq. (\ref{Lescalar}).  So, we can say that the mapping $$\mbox{noncommutativity} \quad \Rightarrow \quad \mbox{commutativity}$$ does not apply here.  In fact we have the mapping $$(x+\theta)-\mbox{space} \quad\Rightarrow \quad\mbox{x-space}$$ zeroing the parameter $\lambda$ and not the NC coordinate, of course.  In other words, this mapping can be obtained carrying out a dimensional reduction in this $(x+\theta)$-space.

\section{Conclusions}

The idea of noncommutativity brings hope to the elimination of divergences that plague QFT.  Snyder was the first one who published a way to deal with these ideas but Yang showed that the divergences were still there.  This result provoke an hibernation of Snyder work in particular and of the noncommutativity concepts in general for more than forty years.  Calculations concerning string theory algebra demonstrated that nature can be NC.  Since string theory is one of the candidates to unify QM with general relativity, noncommutativity concepts were vigorously reborn through a huge and dynamical literature.

To describe some NC aspects, one way that can be used is to analyze noncommutativity through the Moyal-Weyl product where the standard product of two or more fields is substituted by a star product.  In this case, the mathematical consistency of this star product is guaranteed because the NC parameter is constant (\cite{jhep} and references therein.  However, there are NC versions, not using the Moyal-Weyl product, where the NC parameter is not a constant.  We can also introduce noncommutativity through the so-called Bopp shift.  These ones are the most popular realizations of the NC concepts.

In this paper we work with an alternative and ingenuous formulation of NC theory developed recently, motivated by the ideas that in the early Universe, the spacetime may be NC.  This, by the way, is one of the main reasons that fueled NC research.  In this formulation (DFR) the NC parameter is a coordinate of spacetime.  So we have, in a D=4 Minkowski spacetime, for example, a NC space with six $\theta^{\mu\nu}$ coordinates, namely $D(D-1)/2$ NC coordinates.  In this NC spacetime, the $\theta^{\mu\nu}$-coordinate has its conjugated momenta $\pi_{\mu\nu}$ (DFRA). Besides, to work with a ten dimensional NC spacetime can disclose new physics beyond the Standard Model..

On the other hand, to work with NCQM we need only three space coordinates and consequently the NC sector has three coordinates also.  These both sets combined are the so-called DFRA space has been developed through these last few years using the standard concepts of QM and constructing an extension of the Hilbert space.

Here we show that the alternative Feynmann vision for QM can be treated in this DFRA space.  We use this new formalism to quantize the NC free particle and the NC harmonic oscillator.  After that we also constructed the generating functional and the effective action to originate 1PI diagrams.  Finally, we encompassed self-interaction scalar theory $\phi^{n}\,(n\geq 3)$ setting the basis to obtain the correction terms of the perturbation theory of both the propagator and vertex of this model.

As a final remark we would like to say that the formalism developed here is very unusual concerning the fact that it was totally constructed within a NC space.  We did not introduce (by hand) any NC parameter such that its elimination recover the commutative theory.  To recover its commutative behavior, we have to perform a dimensional reduction in this $(x+\theta)$-space which has nine dimensions.  Since some thermodynamics were considered and some objects like transition amplitude and partition function with NC coordinates were calculated, we believe that the construction of a NC thermofield dynamics can be the next move in this direction.  Some cosmological models like black holes and wormholes can be a target for this analysis.  The idea would be, instead of introducing noncommutativity in such models existing in commutative spacetime, to construct black holes and wormholes in a NC spacetime like the one discussed here.


\begin{thebibliography}{30}
\bibitem{NCQM} In Refs.   \cite{Amorim1} and     \cite{Amorim4} it is possible to
find a large amount of references concerning NC quantum mechanics.
\end{thebibliography}
\end{document}